\documentclass[fleqn,12pt,twoside]{article}
\usepackage{espcrc1,graphics}

\usepackage{graphicx}
\usepackage[figuresright]{rotating}

\newcommand{\AmS}{{\protect\the\textfont2
  A\kern-.1667em\lower.5ex\hbox{M}\kern-.125emS}}

\title{$\rho(770)^0$, K$^*(892)^0$ and f$_{0}(980)$ Production in
        Au-Au and pp Collisions at $\sqrt{s_{NN}}$ = 200 GeV}

\author{P. Fachini\address{Brookhaven National
        Laboratory, Bldg. 510A, Upton, NY 11973, USA} for the STAR Collaboration\thanks{For
        the full author list and acknowledgements, see Appendix ``Collaborations'' of this
        volume.}}%

\begin{document}

\maketitle

\begin{abstract}
Preliminary results on $\rho(770)^0 \rightarrow \pi^{+}\pi^{-}$,
K$^{*}(892)^{0} \rightarrow \pi$K and $f_{0}(980) \rightarrow
\pi^{+}\pi^{-}$ production using the mixed-event technique are
presented. The measurements are performed at mid-rapidity by the
STAR detector in $\sqrt{s_{NN}}$= 200 GeV Au-Au and pp
interactions at RHIC. The results are compared to different
measurements at various energies.
\end{abstract}

\section{Introduction}

The measurement of resonances that have lifetimes smaller than, or
comparable to, the lifetime of the dense matter produced in
relativistic heavy-ion collisions provides an important tool for
studying the collision dynamics. Resonances that decay before
kinetic freeze-out may not be reconstructed due to the
re-scattering of the daughter particles. Therefore, short-lived
resonance production may provide information on the time between
chemical and kinetic freeze-out. The measurement of resonances
helps constraining thermal models via their feed-down to stable
hadrons and their ratios to other particles \cite{1}. In addition,
the measurement of $\rho(770)^0 \rightarrow \pi^{+}\pi^{-}$
production can provide information for studying the di-lepton
decay channel, since the $\rho(770)^0 \rightarrow \ell^+\ell^-$
production in the hadronic `cocktail' is currently based on model
calculations \cite{17}. First results on $\rho(770)^0$,
K$^*(892)^0$ and f$_0(980)$ measurements via their hadronic decay
channel in Au-Au and pp collisions at $\sqrt{s_{NN}}$ = 200 GeV
using the STAR detector at RHIC are presented. The results are
compared to the measurements in Au-Au, pp, $\bar{\textrm{p}}$p and
e$^{+}$e$^{-}$ interactions at various energies.

\section{Data Analysis and Results}

The main STAR detector consists of a large Time Projection Chamber
(TPC) \cite{5} placed inside an uniform solenoidal magnetic field
of 0.5 T. The TPC provides the measurement of charged particles.
About 2M Au-Au minimum bias events and about 4.7M pp events at
$\sqrt{s_{NN}}$ = 200 GeV are used in this analysis. The
$\rho(770)^0$, K$^*(892)^0$ and f$_0(980)$ measurements follow our
previous K$^*(892)^0$ measurement at $\sqrt{s_{NN}}$= 130 GeV
\cite{6}. Due to limited statistics, the term K$^{*0}$ in this
analysis refers to the average of K$^{*0}$ and
$\overline{\textrm{K}^{*0}}$ unless specified.

\begin{figure}[htb]
\vspace{-0.2in}
\begin{minipage}[t]{80mm}
\includegraphics[height=14pc,width=18pc]{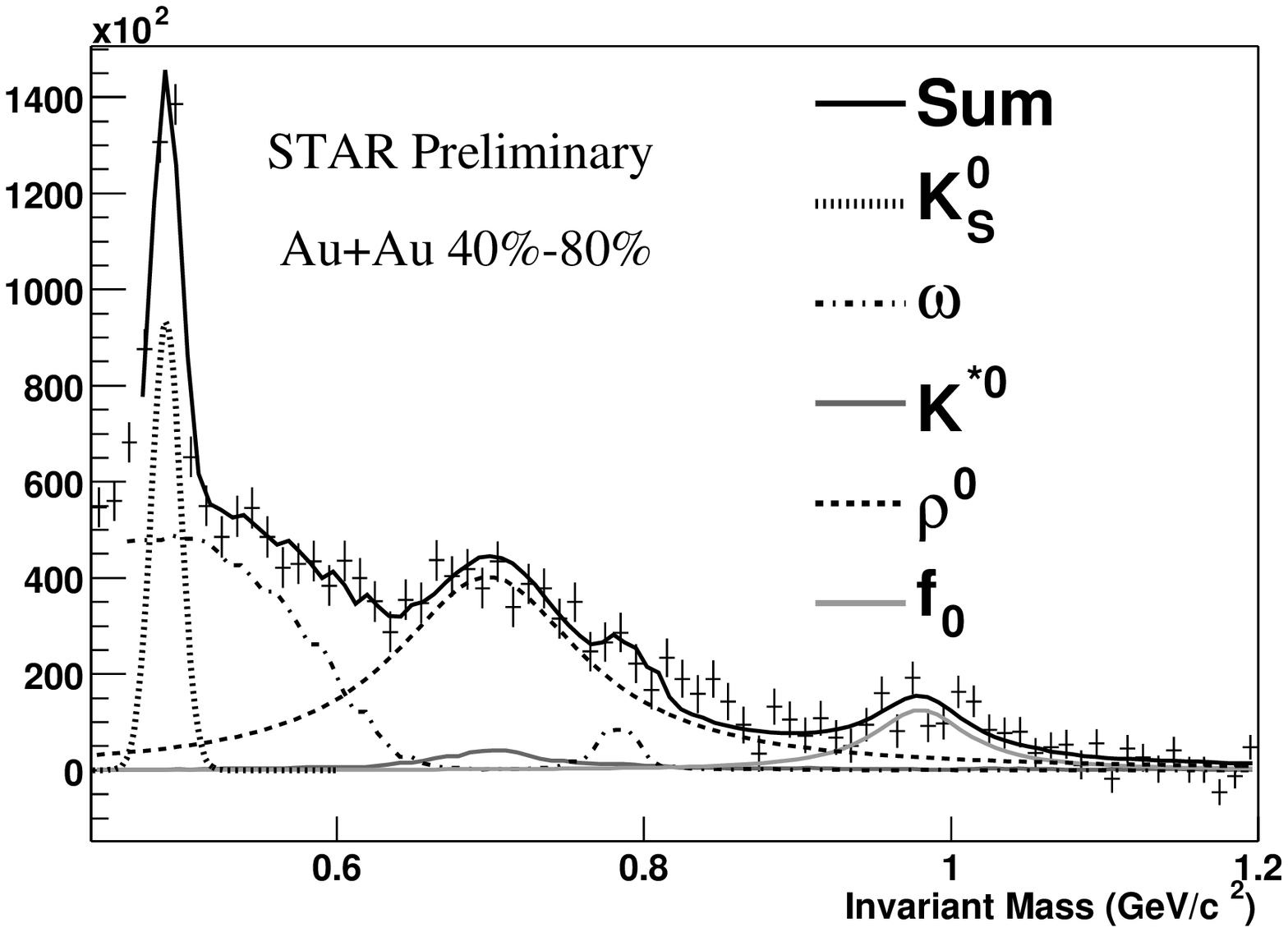}
\end{minipage}
\hspace{\fill}
\begin{minipage}[t]{85mm}
\includegraphics[height=14pc,width=18pc]{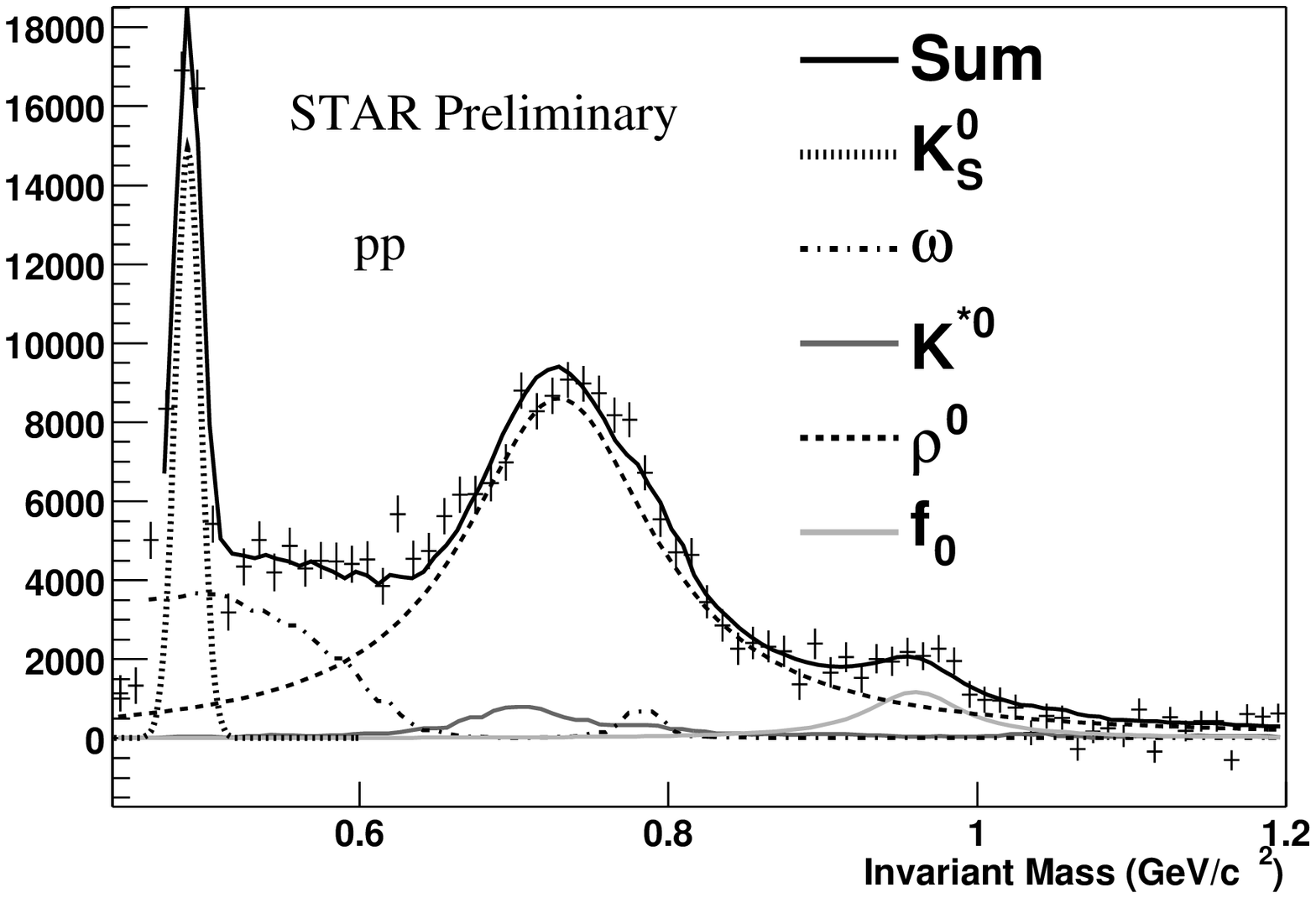}
\end{minipage}
\caption{The $\pi^+\pi^-$ invariant mass distributions after
background subtraction for the 40$\%$ to 80$\%$ of the hadronic
Au-Au cross section (left) and for pp interactions (right).}
\label{Cocktail} \vspace{-0.2in}
\end{figure}

The $\pi^+\pi^-$ invariant mass distributions after background
subtraction for the 40$\%$ to 80$\%$ of the hadronic Au-Au cross
section and for pp interactions are shown in Fig. \ref{Cocktail}.
The K$_S^0$ is fit to a gaussian (dotted line), the $\omega$ shape
(dash-dotted line) is obtained from the HIJING event generator,
the K$^{*0}$ shape (dark grey line) is also obtained from HIJING
with the kaon being misidentified as a pion, and both the $\rho^0$
(dashed line) and the f$_0$ (light grey line) are fit to a
Breit-Wigner function with a fixed width of 150 MeV and 75 MeV,
respectively. In the case of the $\rho^0$ and f$_0$ line shapes,
we are investigating distortions due to final state interactions
and production via re-scattering.
The $\omega$ signal is a free parameter in the fit and the
K$^{*0}$ signal is fixed according to the K$^{*}(892)^{0}
\rightarrow \pi$K measurement. The solid black line corresponds to
sum of all these correlations in the $\pi^+\pi^-$ channel. Since
the identification of the daughter particles from the resonance
decay is obtained by the energy loss in the gas of the TPC,
particle misidentification increases with p$_T$. As a consequence,
the K$^{*0}$ contamination in the $\pi^+\pi^-$ channel limits the
measured p$_T$ range of the $\rho^0$.


\begin{figure}[htb]
\vspace{-0.3in} \begin{minipage}[t]{80mm}
\includegraphics[height=13pc,width=18pc]{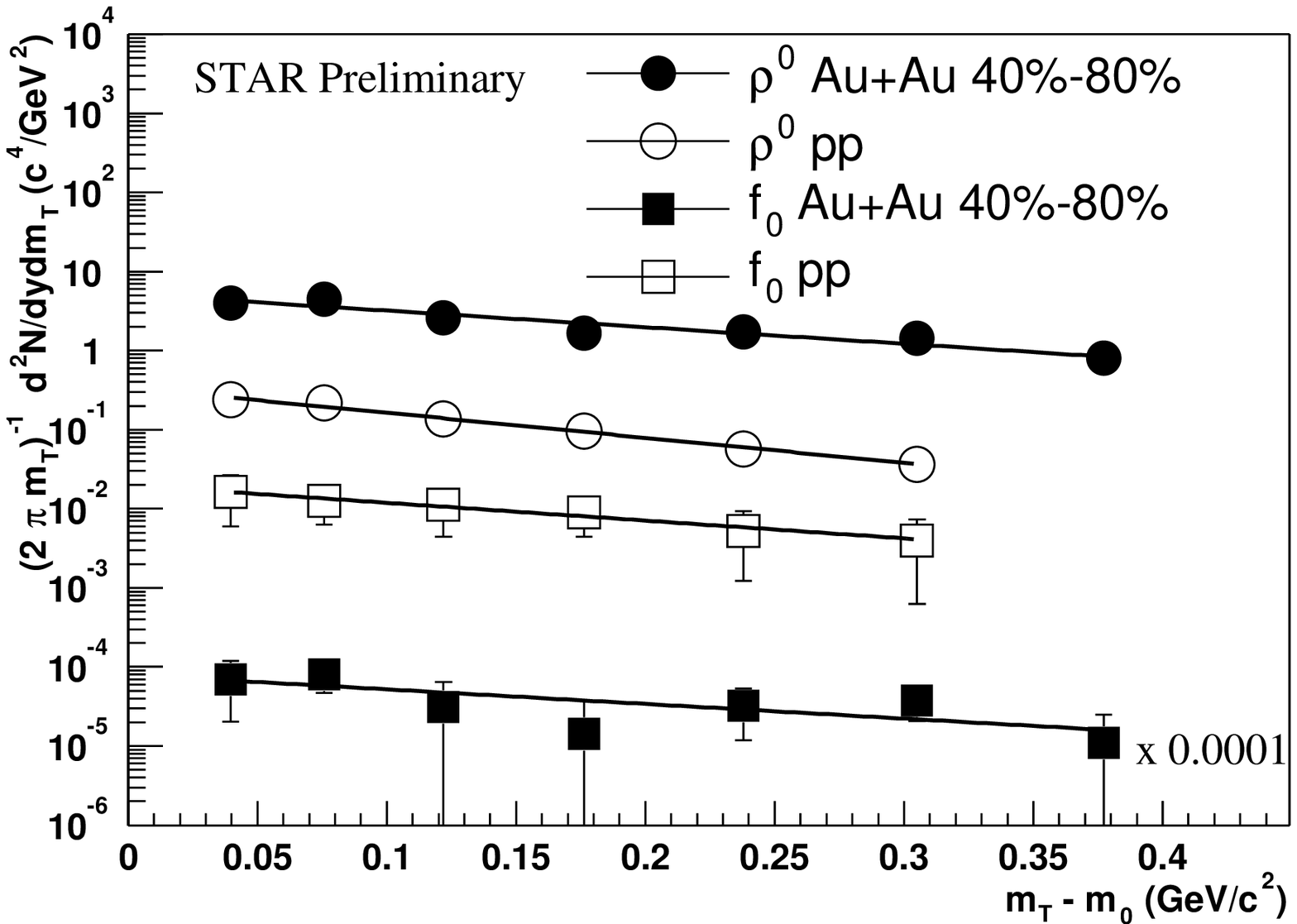}
\end{minipage}
\hspace{\fill}
\begin{minipage}[t]{85mm}
\includegraphics[height=13pc,width=18pc]{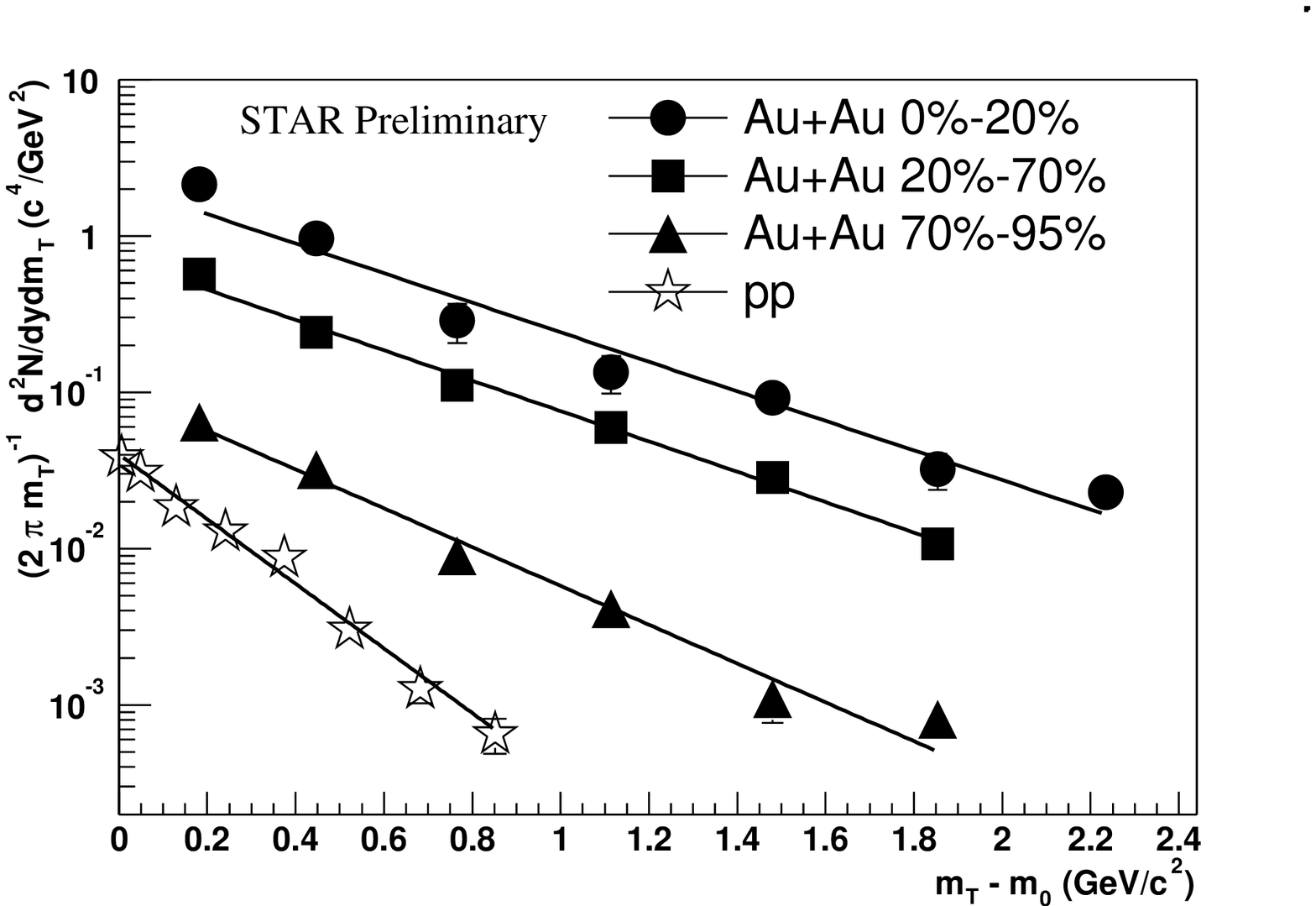}
\end{minipage}
\caption{m$_T$-distributions at mid-rapidity ($|y| <$ 0.5) for
$\rho^0$, f$_0$ (left) and (K$^{*0} +
\overline{\textrm{K}^{*0}})/2$ (right) from Au-Au and pp
interactions. The errors shown are statistical only.}
\label{Spectra} \vspace{-0.2in}
\end{figure}

In the case of the K$\pi$ invariant mass distribution after
background subtraction, a combination of a linear background and a
Breit-Wigner function are used to fit the K$^{*0}$ signal with the
nominal resonance width and mass, as described in \cite{6}.

The $\rho^0$, f$_0$ and (K$^{*0} + \overline{\textrm{K}^{*0}})/2$
m$_T$-distributions at mid-rapidity ($|y| <$ 0.5) from both Au-Au
and pp interactions are shown in Fig. \ref{Spectra}. An
exponential fit is used to extract the yields per unit of rapidity
and the inverse slopes. The systematic uncertainty in the K$^{*0}$
dN/dy is estimated to be 25$\%$ for Au-Au and 10$\%$ for pp
collisions due to detector effects and the uncertainty in the
background determination. The systematic uncertainty in the
$\rho^0$ and f$_0$ dN/dy is estimated to be 30$\%$ and 50$\%$,
respectively for both pp and Au-Au collisions.
The measurements from pp interactions are
not corrected for trigger bias and vertex finding efficiency.
However, these corrections may cancel in the particle ratios.

\begin{figure}[htb]
\vspace{-0.05in}
\begin{minipage}[t]{85mm}
\includegraphics[height=14pc,width=18pc]{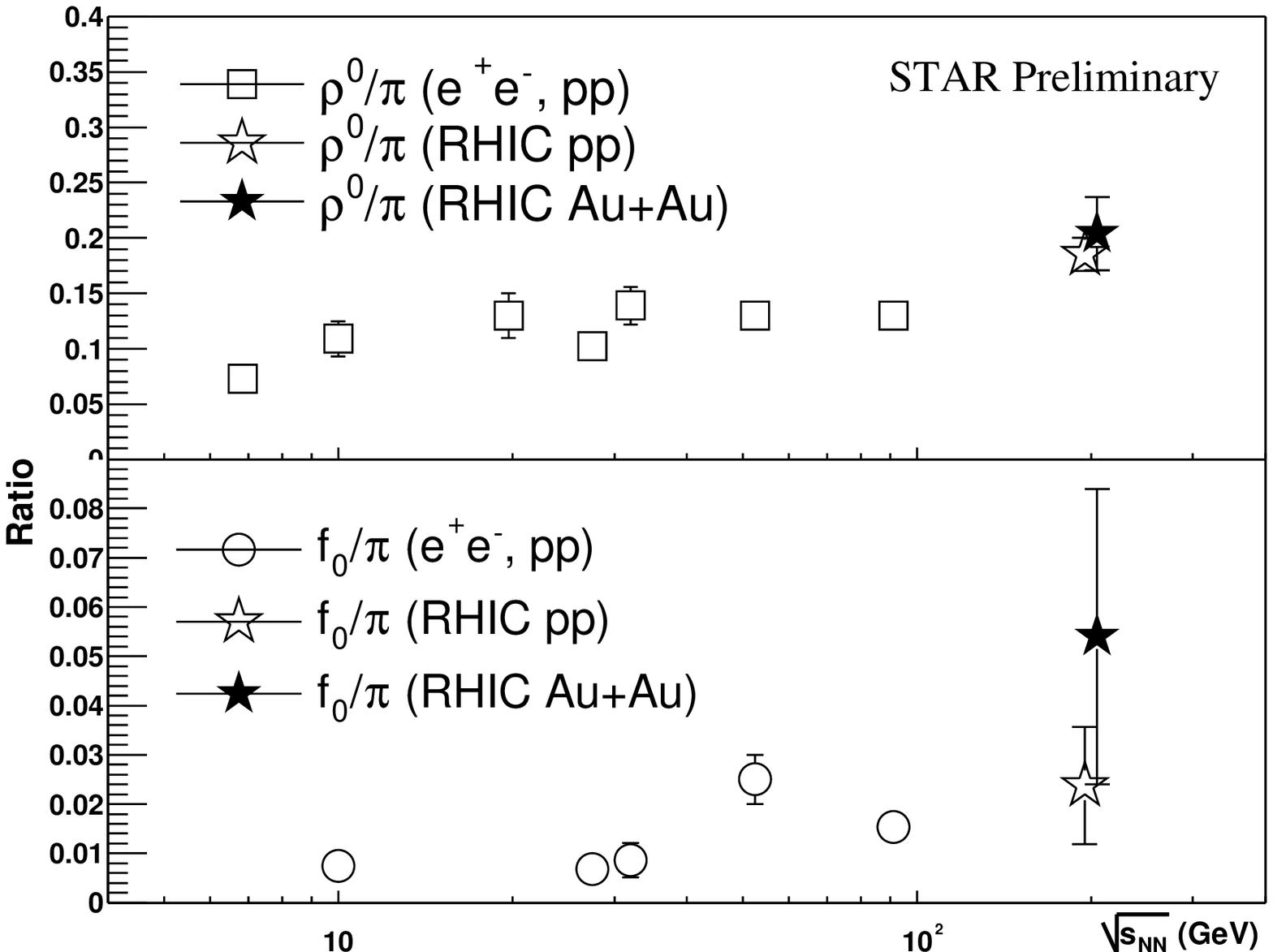}
\end{minipage}
\hspace{\fill}
\begin{minipage}[t]{80mm}
\includegraphics[height=14pc,width=18pc]{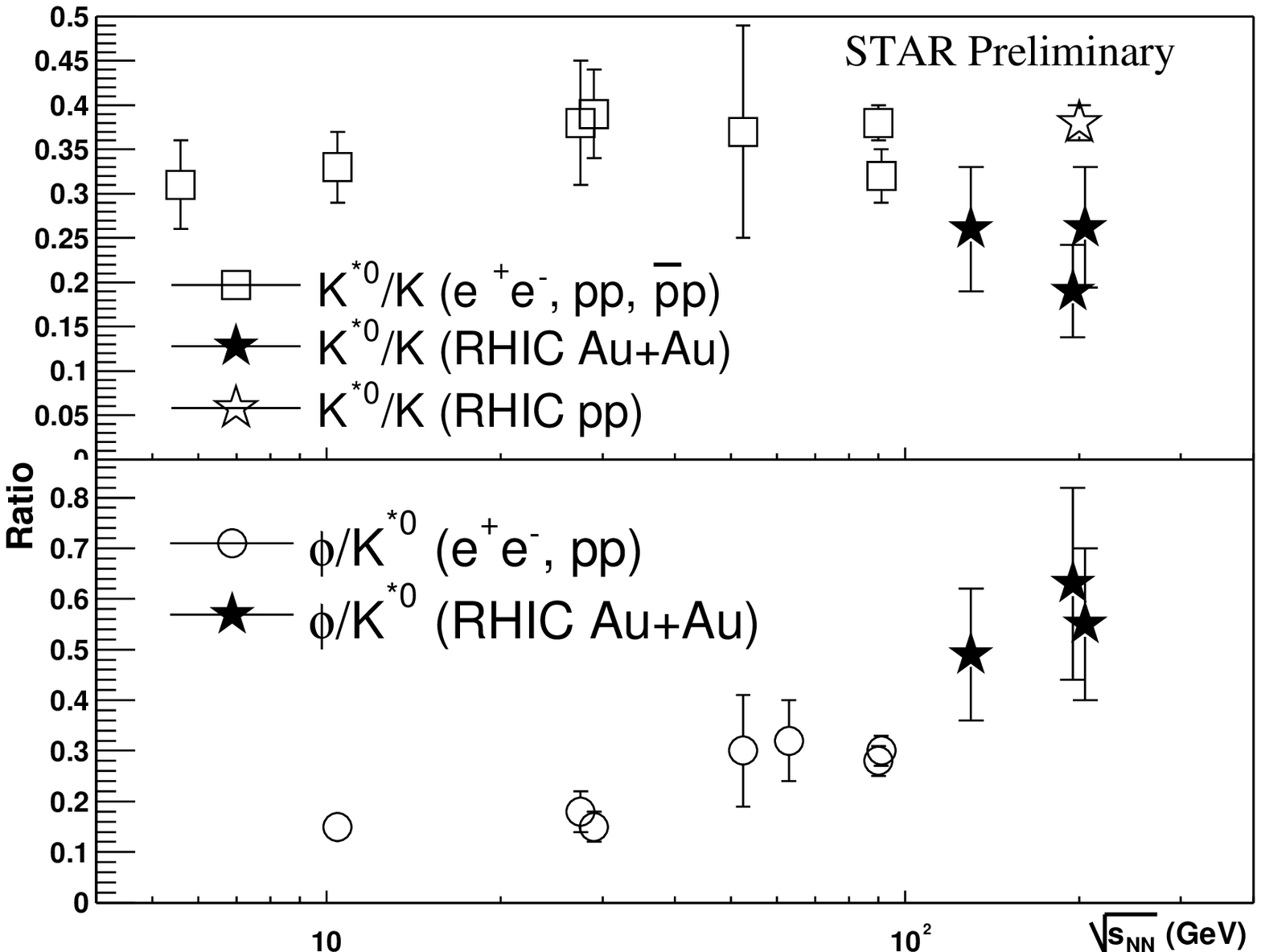}
\end{minipage}
\caption{$\rho^0/\pi$, f$_0/\pi$ (left), K$^{*0}$/K and
$\phi$/K$^{*0}$ (right) ratios as a function of beam energy. The
$\rho^0/\pi$ and f$_0/\pi$ ratios from Au-Au collisions correspond
to 40-80$\%$ of the hadronic cross-section. The K$^{*0}$/K and
K$^{*0}/\phi$ ratios from Au-Au collisions at $\sqrt{s_{NN}}$= 200
GeV correspond to 0-20$\%$ and 20-70$\%$ of the hadronic
cross-section. The ratios are from measurements in e$^+$e$^-$
collisions at 10.45 GeV \cite{7}, 29 GeV \cite{8}, 90 GeV \cite{9}
and 91 GeV \cite{10} beam energies, $\bar{\textrm{p}}$p at 5.6 GeV
\cite{11} and pp at 6.8 GeV \cite{15}, 19.7 GeV \cite{16}, 27.5
GeV \cite{12}, 52.5 GeV \cite{13} and 63 GeV \cite{14}. The errors
on the K$^{*0}$/K and $\phi$/K$^{*0}$ ratios at $\sqrt{s_{NN}}$=
200 GeV correspond to the quadratic sum of the statistical and
systematic errors. The errors on the other ratios at
$\sqrt{s_{NN}}$= 200 GeV are statistical only.} \label{Ratios}
\vspace{-0.2in}
\end{figure}

In the $\rho^0$ analysis, we follow previous e$^+$e$^-$ and pp
measurements that do not select exclusively on the $\ell$=1
$\pi^+\pi^-$ channel. Figure \ref{Ratios} depicts the $\rho^0/\pi$
and f$_0/\pi$ ratios as a function of beam energy for different
colliding systems. Both $\rho^0$ and f$_0$ production measured at
RHIC seem to follow the trend of previous measurements.

Figure \ref{Ratios} also shows the K$^{*0}$/K and $\phi$/K$^{*0}$
ratios measured in different colliding systems at various
energies. The K$^{*0}$/K ratio is interesting because K$^{*0}$ and
K have similar quark content and differ mainly in mass and spin.
From Fig. \ref{Ratios}, the K$^{*0}$/K ratio from $\sqrt{s_{NN}}$=
200 GeV Au-Au collisions is lower than the pp measurement at the
same energy by a factor of 2.

The $\phi$/K$^{*0}$ ratio measures the strangeness suppression in
nearly ideal conditions since $\Delta$S = 1, since strangeness is
hidden in the $\phi$, and there is only a small mass difference.
Fig. \ref{Ratios} shows an increase of the ratio $\phi$/K$^{*0}$
measured in Au-Au collisions compared to the measurements in pp
and e$^+$e$^-$ at lower energies. However, this increase may not
be solely related to strangeness suppression due to additional
effects on short lived resonances in heavy-ion collisions.

The K$^{*0}$ lifetime is short (c$\tau$ = 4 fm) and comparable to
the time scale of the evolution of the system in relativistic
heavy-ion collisions. If the K$^{*0}$ decay between chemical and
kinetic freeze-out, the daughters from the decay may re-scatter
and the K$^{*0}$ will not be reconstructed. On the other hand,
elastic interactions are still effective after chemical freeze-out
\cite{4} and may regenerate the K$^{*0}$ until kinetic freeze-out.
Assuming that the difference in the K$^{*0}$/K ratio measured in
$\sqrt{s_{NN}}$= 200 GeV Au-Au and pp collisions is due to the
K$^{*0}$ survival probability, our measurement is consistent with
a time of only a few fm between chemical and kinetic freeze-out
(scenario \cite{2}). Hence, our measurement at RHIC is consistent
with either a sudden freeze-out with no K$^{*0}$ regeneration or a
long time scenario ($\sim$20 fm \cite{3}) with significant
K$^{*0}$ regeneration.

\section{Conclusions}
Preliminary results on $\rho(770)^0$, K$^{*}(892)^{0}$ and
f$_0$(980) production measured at mid-rapidity by the STAR
detector in $\sqrt{s_{NN}}$= 200 GeV hadronic Au-Au and pp
interactions at RHIC were presented. The data show significant
production of these short-lived resonances. The K$^{*0}$
production at RHIC rules out a long expansion time between
chemical and kinetic freeze-out unless there is significant
K$^{*0}$ regeneration, due to elastic interactions after chemical
freeze-out. Finally, the study of short-lived resonances may
provide important information on the collision dynamics.

\end{document}